\documentclass[a4paper,11pt]{article}
\usepackage{pos}
\usepackage{slashed}

\title{Renormalization of Karsten-Wilczek Quarks on a Staggered Background}

\author*[a]{Daniel A. Godzieba}
\author[b]{Szabolcs Borsanyi}
\author[a,b]{Zoltan Fodor}
\author[a,c]{Paolo Parotto}
\author[b]{R\'eka A. Vig}
\author[b]{Chik Him Wong}

\affiliation[a]{Pennsylvania State University, Department of Physics, University Park, PA 16802, USA}

\affiliation[b]{University of Wuppertal, Department of Physics, Wuppertal D-42119, Germany}

\affiliation[c]{Dipartimento di Fisica, Universit\`a di Torino and INFN Torino, Via P. Giuria 1, I-10125 Torino, Italy}

\emailAdd{dag5611@psu.edu}
\emailAdd{borsanyi@uni-wuppertal.de}
\emailAdd{zxf5098@psu.edu}
\emailAdd{paolo.parotto@gmail.com}
\emailAdd{tajhajlito@gmail.com}
\emailAdd{cwong@uni-wuppertal.de}

\abstract{The Karsten-Wilczek action is a formulation of minimally doubled fermions on the lattice. It explicitly breaks hypercubic symmetry and introduces three counterterms with respective bare parameters. We present a tuning of the bare parameters of the Karsten-Wilczek action on staggered configurations at the physical point. We study the magnitude of the taste-splitting as a function of the lattice spacing.}

\FullConference{The 40th International Symposium on Lattice Field Theory (Lattice 2023)\\
July 31st - August 4th, 2023\\
Fermi National Accelerator Laboratory\\}

\begin{document}
\maketitle

\section{Introduction}

The Karsten-Wilczek (KW) action \cite{Karsten:1981gd,Wilczek:1987kw} is the simplest
implementation of so-called minimally doubled fermions in lattice field theory.
The action eliminates most of the spurious fermionic degrees
of freedom known as ``doublers.'' It reduces the number of doublers from
fifteen to one, leaving two mass-degenerate quark species, while explicitly
preserving an ultra-local chiral symmetry. In the case of degenerate up
and down quarks this action allows the study of the chiral transition
without a rooting, which one is forced to employ with staggered quarks.

However, minimally doubled fermions come at a price. The action breaks the 
hypercubic symmetry of the lattice and introduces three counterterms to the 
na\"ive theory \cite{Weber:2015oqf}. While the KW action is highly appealing 
for the aforementioned features, renormalization constitutes a
multidimensional tuning problem for the selection of appropriate values for
the bare parameters. 

We present a mixed action study of a method of tuning these bare parameters 
nonperturbatively, exploring the hierarchy of the tuning parameters and how 
accurately one needs to tune. We take measurements with the KW action on stored 
gauge configurations computed with the 4-stout staggered fermion action at the 
physical point. We conclude with an exploration of the scaling behavior of the 
mass-splitting of the ground states of two mesonic channels with tuned 
parameters.


\section{The Karsten-Wilczek action} \label{KW}

The Karsten-Wilczek action \cite{Karsten:1981gd,Wilczek:1987kw} \begin{equation}
S_F^{KW} = S_F^N + \sum_x \sum_{j=1}^3 \bar{\psi}(x) \frac{i\zeta}{2} \gamma^\alpha \left( 2\psi(x) - U_j(x) \psi(x+\hat{j}) - U_j^\dagger(x) \psi(x - \hat{j}) \right), \label{eq:KW_action}
\end{equation} where $\zeta$ is the Wilczek parameter, $U_{\mu}(x)$ is the link variable at site $x$ in the $\mu$ direction, and $\alpha$ is an arbitrary spacetime direction, adds to the na\"ive lattice fermion action \begin{equation}
S_F^N=\sum_x \sum_{\mu=0}^3 
\bar\psi(x) \gamma_\mu \frac12 \left[U_\mu(x)
\psi(x+\hat\mu) - U^+_\mu(x-\hat\mu)\psi(x-\hat\mu)
\right] + m_0 \sum_x \bar\psi(x)\psi(x)
\end{equation} additional terms which break hypercubic symmetry. These terms
are only added along the spacetime directions where $\mu\neq\alpha$, making
$\alpha$ a special axis and explicitly breaking the hypercubic symmetry of
$S_F^N$. In this work, we take $\alpha = 0$, the time direction.
The action requires the following three counterterms, for which
one-loop perturbative results exist \cite{Capitani:2010nn}
\begin{align} 
S^{3f}
&= c \sum_x \bar{\psi}(x) i\gamma^\alpha \psi(x), \\
S^{4f} &= d \sum_x \bar{\psi}(x)\frac{1}{2}\gamma^\alpha \left( U_\alpha(x) \psi(x+\hat{\alpha}) - U_\alpha^\dagger(x) \psi(x - \hat{\alpha}) \right), \\
S^{4g} &= d_G \sum_x \sum_{\mu \neq \alpha} {\rm Re}\, {\rm Tr}(1 - \mathcal{P}_{\mu\alpha}(x))\,.
\end{align} 
Here $\mathcal{P}_{\mu\nu}(x)$ is the plaquette at $x$ along the $\mu$ and
$\nu$ directions. $S^{3f}$ and $S^{4f}$ are fermionic counterterms of dimension
3 and 4 respectively. The gluonic counterterm $S^{4g}$ plays no role in our
mixed action study. It will be relevant for dynamical KW
simulations.

These terms are manifestations of the anisotropy of the lattice breaking
hypercubic symmetry \cite{Weber:2015oqf}. The counterterms $S^{4f}$ and
$S^{4g}$ act as anisotropy parameters of the fermion and gauge terms,
respectively. To emphasize the role of $S^{4f}$, we express its bare parameter $d$ in terms of the anisotropy $\xi_0=1+d$ in our numerical work.
Following the preferred axis $\alpha=0$, we target a renormalization anisotropy
$\xi_f=1$ and find the bare anisotropy $\xi_0$ that restores isotropy.

The multidimensional nonpertrubative tuning of ($c$, $\xi_0$)
can be made tractable with certain considerations. Boosted
perturbation theory estimates of the nonperturbative parameters and quenched
simulations using the KW action indicate that the anisotropy parameter has mild
effects compared to those of $c$ and that $c$ can be considered almost
independent of $\xi_0$ \cite{Weber:2015oqf}. We quantify this statement
and explore a method for tuning the relevant parameter $c$ and the bare
anisotropy $\xi_0$. Having an efficient method will be critical eventually when
tuning dynamical KW simulations \cite{Reka}. 

We use the KW action with $\alpha=0$, $\zeta={+}1$. We use the $w_0$ scale
setting to convert from lattice units into physical units \cite{BMW:2012hcm}.
The 4-stout action's parameters were introduced in Ref.~\cite{Bellwied:2015lba}.


\section{Nonperturbative tuning of bare parameters} \label{taste_splitting}

A suitable method for tuning the bare parameters of the KW action nonperturbatively uses the existence of oscillating contributions to the correlation functions\begin{equation} C_\mu(x,y) \sim \left\langle \bar{\psi}(x) \gamma_\mu \psi(x) \bar{\psi}(y) \gamma_\mu \psi(y) \right\rangle \end{equation} of some KW fermions. These oscillations are related to fermion doubling \cite{Weber:2015oqf}. Importantly, their frequency is sensitive to the $c$ parameter. Fermion partners can be identified by the spin-taste structure of the KW action \cite{Weber:2023kth}. A relevant fermion pair is the $\gamma_0$ and $\gamma_5$ channels. The correlators for $\gamma_0$ and $\gamma_5$ are taken \textit{parallel} to the preferred axis of the KW action, $\alpha=0$ in this case. The correlator of $\gamma_5$ \textit{perpendicular} to $\alpha=0$ in a spatial direction is needed for tuning $\xi_0$. The renormalized fermionic anisotropy is defined as the ratio of the perpendicular (spatial) mass of $\gamma_5$ to the parallel (temporal) mass: $\xi_f = m_\perp/m_\parallel$.

\begin{figure}
 \centering 
\includegraphics[trim=5 0 40 15, clip, width=.45\textwidth]{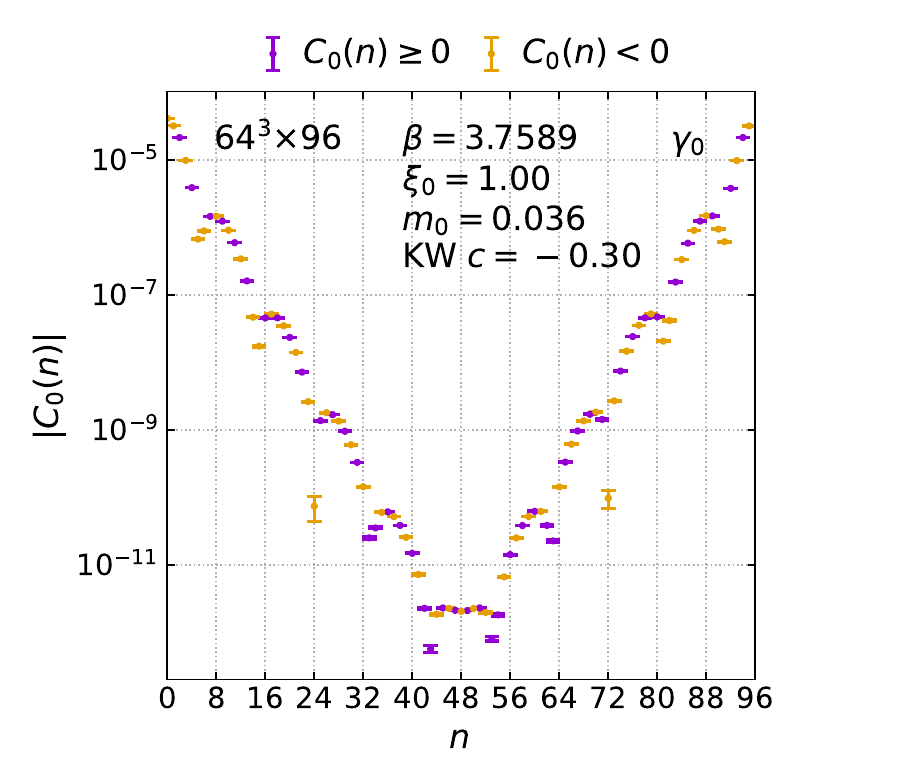} 
\includegraphics[trim=5 0 40 15, clip, width=.45\textwidth]{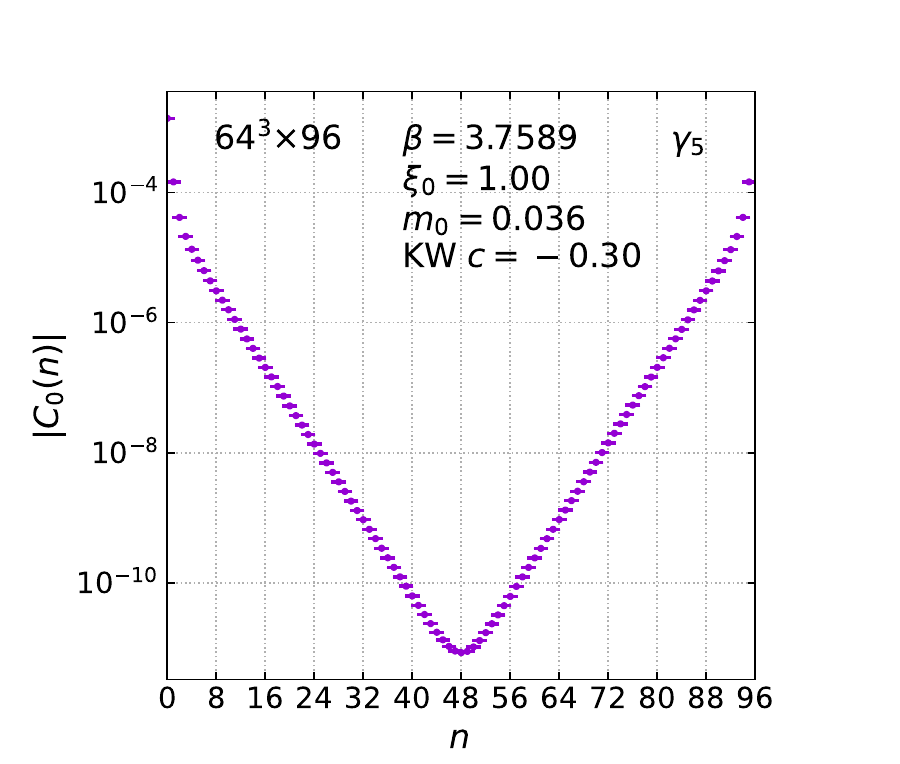}
\caption{Correlators for the $\gamma_0$ (left) and $\gamma_5$ (right) fermion
channels for the Karsten-Wilczek action. The frequency of the oscillating
$\gamma_0$ correlator depends on the $c$ parameter of the action. A beat
oscillation on top of the $(-1)^n$ oscillation (visible in the left plot)
occurs when the $c$ parameter is detuned. \label{fig:correlators} }
\end{figure}

The parallel correlator for $\gamma_0$ exhibits oscillations while that of $\gamma_5$ does not. The tuning criterion for the $c$ parameter is where the frequency spectrum of the oscillations of the $\gamma_0$ channel recovers its tree-level form \cite{Weber:2015oqf}. At the tuned $c$, the oscillation of the $\gamma_0$ correlator is described by $(-1)^n$, where $n$ is the position along the temporal extent of the lattice. The frequency of the oscillation at a given $c$ value can be described by $\omega = \omega_c + \pi$, where $\pi$ is the frequency of the rapid oscillations at tuned $c$ and $\omega_c$ is a beat frequency that appears in the rapid oscillations when $c$ is detuned. Thus, the tuning criterion is equivalently stated as where $\omega_c = 0$ and the beat vanishes. Fig. \ref{fig:correlators} shows an example of correlators in the $\gamma_0$ and $\gamma_5$ channels at detuned $c$ where both the rapid oscillation and the beat oscillation can be observed for $\gamma_0$. When we average over the symmetric halves of the $\gamma_0$ correlator $C_0(x,y)$ and eliminate the rapid oscillation with a factor of $(-1)^n$,
\begin{equation}
C(n) = (-1)^n\frac{1}{2}(C_0(0,n) + C_0(0,N_t-n)) \quad \text{for } 0 \leq n \leq N_t/2,
\end{equation} where $N_t$ is the temporal extent of the lattice, the correlator is well described by the model \begin{equation}
C(n) \approx A \cosh(m(n-N_t/2)) \cos(\omega_c n - \phi) 
\end{equation}
where $m$ is the mass of $\gamma_0$ in lattice units and $\phi$ is a phase factor. The beat frequency and the mass decouple in this way. We use various fitting methods for extracting $m$ and $\omega_c$ because of the exotic shape of the correlator. We tune $c$ for a given $\xi_0$ by scanning through $c$ and extrapolating with a linear fit to where $\omega_c = 0$.

\begin{figure}
 \centering 
\includegraphics[width=.42\textwidth]{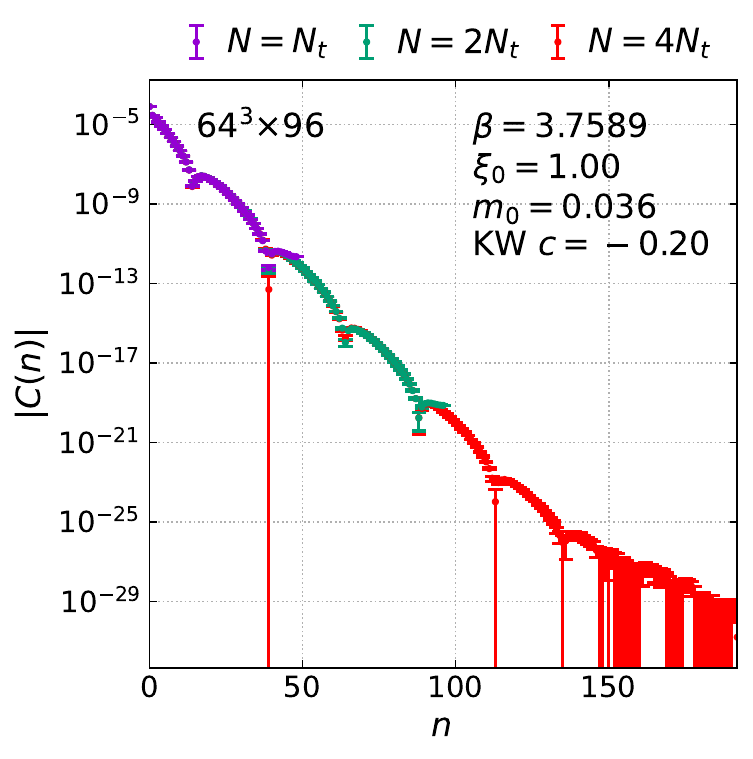}
\includegraphics[width=.42\textwidth]{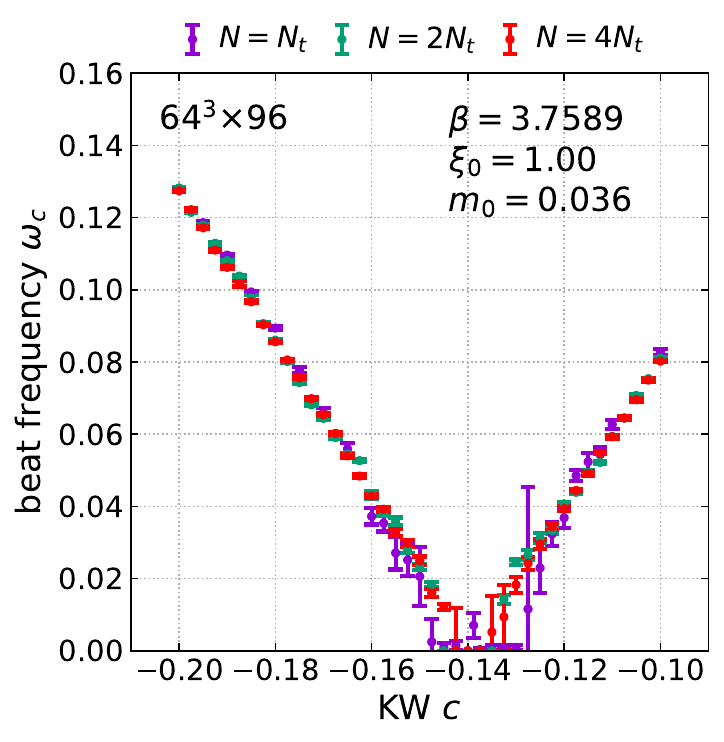}
\caption{(Left) The symmetrized $\gamma_0$ correlator measured with different
amounts of tiling with stored gauge configurations. The proper use of tiling
permits longer correlation lengths to be measured and increases the precision
of the measurement of the beat frequency $\omega_c$. (Right) Measurements of
$\omega_c$ as a function of the $c$ parameter at the same three tilings.
\label{fig:tiling} }
\end{figure}

An immediate difficulty that arises in tuning $c$ is the finite length of the
lattice, which makes fitting the frequency unreliable when the wavelength of
the beat is greater than $N_t$ in the relative vicinity of tuned $c$. This can
be ameliorated through the use of tiling gauge configurations. 
Tiling means the gauge configurations are doubled (or quadrupled) to form a
$N_x^3\times(2 N_t)$ or $N_x^3\times(4 N_t)$ lattice on which the propagator is
studied. This allows to study the propagator in a longer range than that of the
dynamical simulation. Tiling extends the idea of a mixed action study where
the quarks live on an extended lattice without maintaining the long wave-length
gauge fluctuations, which would anyway be irrelevant for the divergent parts of 
the diagrams. Thus, when used properly, tiling the stored gauge configurations
increases the precision with which $\omega_c$ is measured. Fig.
\ref{fig:tiling} shows an example of $C(n)$ at detuned $c$ for three different
amounts of tiling in the left plot, as well as $\omega_c(c)$ around tuned $c$
for each level of tiling for comparison in the right plot. We use 4${\times}$
and occasionally 1${\times}$ tiling throughout this study.

\begin{figure}
 \centering 
\includegraphics[width=.505\textwidth]{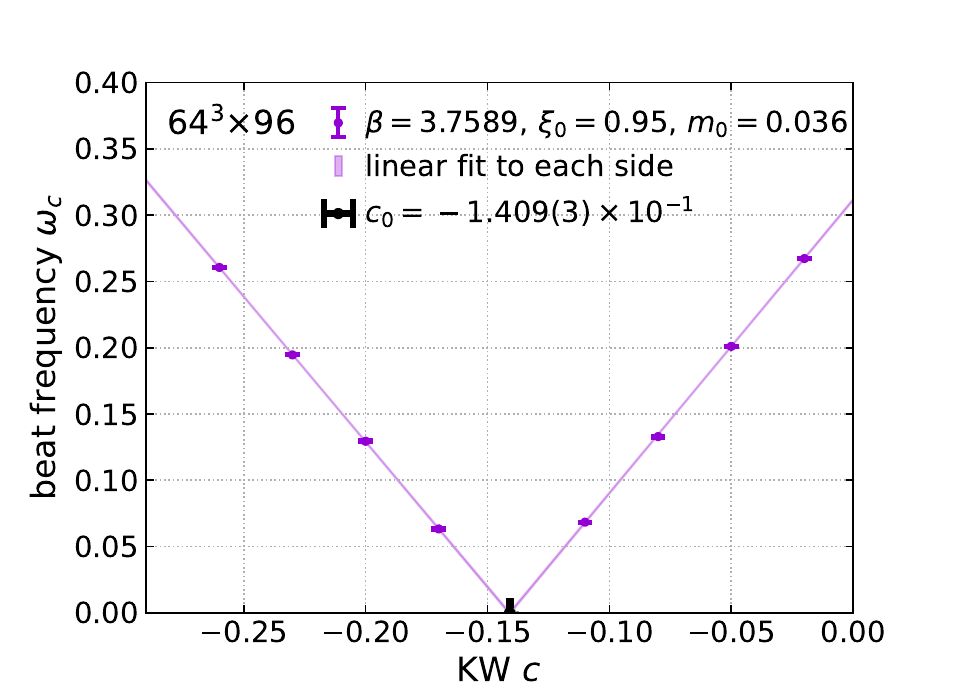} \includegraphics[width=.45\textwidth]{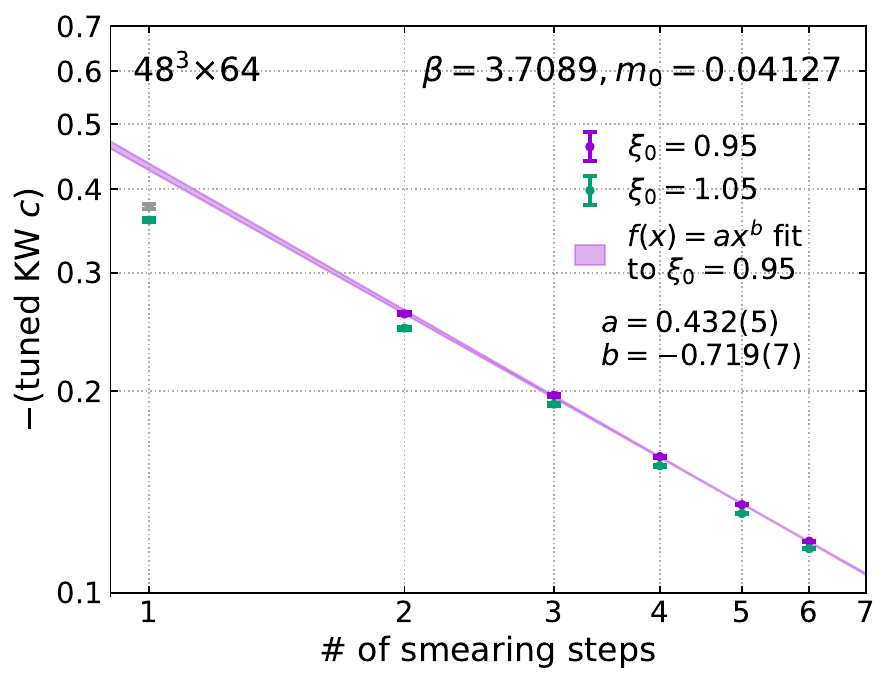}
\caption{(Left) $\omega_c$ as a function of the $c$ parameter. The $c$ parameter is tuned at the value where the beat frequency vanishes. (Right) The variation of the tuned value of $c$ with the number of stout smearing steps applied. We used 4 stout smearing steps throughout the rest of the analysis. \label{fig:c_tuning} }
\end{figure} 

An example of finding the tuned $c$ value is shown in the left plot of 
Fig. \ref{fig:c_tuning}, where the result is combined from linear fits on 
either side of the tuned $c$. An additional parameter for our measurements is 
the number of stout smearing steps we apply. For any value of $\xi_0$, the 
effect of applying more smearing steps is a power law decrease in the magnitude 
of the tuned $c$ value. An example of this is shown in the right plot of 
Fig. \ref{fig:c_tuning}. We consistently use four steps of stout smearing  
throughout this study, the same smearing level that was used in the staggered 
simulation to create the ensembles.

\begin{figure} \centering
\includegraphics[width=.49\textwidth]{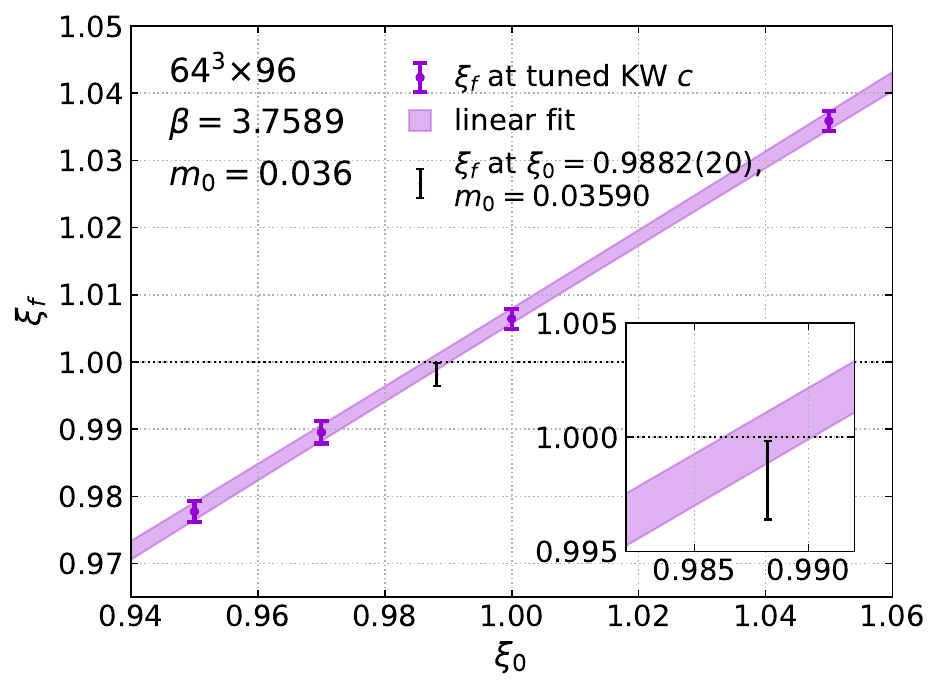}
\includegraphics[width=.5\textwidth]{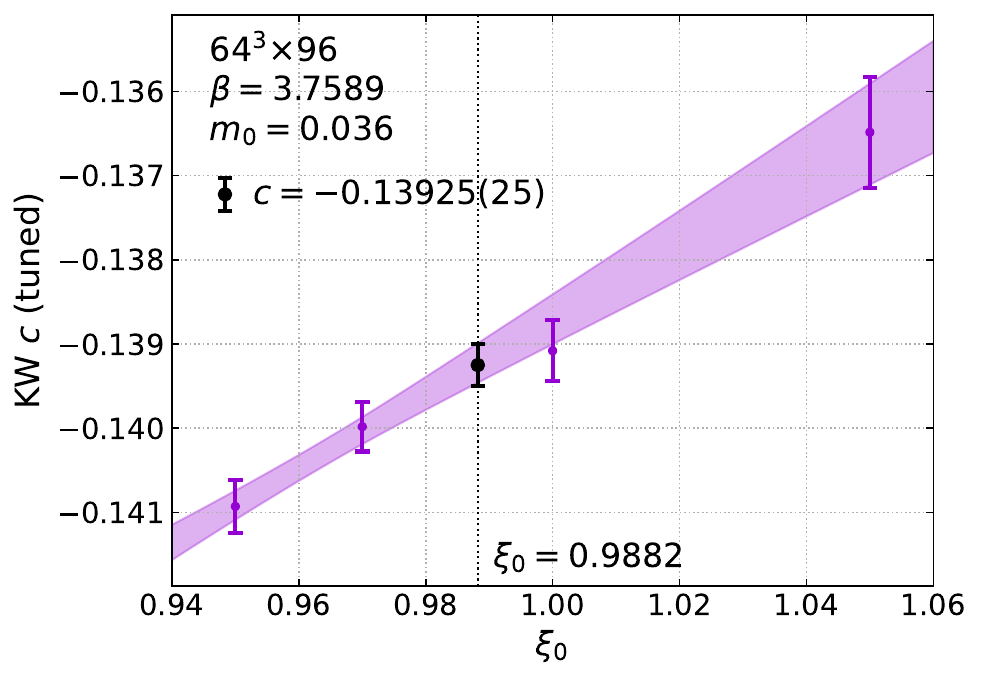} 
\caption{(Left) The renormalized anisotropy $\xi_f$ of the mass of the
pseudoscalar $\gamma_5$ channel as a function the bare anisotropy $\xi_0$.
The $c$ parameter is individually tuned at each $\xi_0$ value. $\xi_f = 1$ was
the tuning criterion for $\xi_0$. (Right) Interpolating to the tuned value of
the $c$ parameter at the tuned value of $\xi_0$. \label{fig:c_xi_tuning} }
\end{figure}

In this way, we tune $c$ for any value of $\xi_0$. $\xi_0$ can then be tuned
according to the desired renormalized anisotropy $\xi_f$. Since we perform
measurements on isotropic staggered configurations, we tune by interpolating
$\xi_f(\xi_0)$ at tuned $c$ to $\xi_f(\xi_0)=1$. An example of this is shown in
the left plot of Fig. \ref{fig:c_xi_tuning}. 
One could proceed to find tuned $c$ at tuned $\xi_0$ using either
a final scan in $c$ at tuned $\xi_0$, or an interpolation of $c(\xi_0)$ to the 
tuned $\xi_0$. We use the latter, keeping in mind the computational efficiency 
of the method. The right plot in Fig. \ref{fig:c_xi_tuning} shows an
example of this interpolation. The question naturally arises as to how 
accurately one needs to tune $c$ and $\xi_0$, or how stable the result is to 
slight mistunings. Thankfully, the results are quite stable, as the $\xi_f(c)$ 
function has vanishing derivative (a maximum) near the tuned value.

\begin{figure}
 \centering 
\includegraphics[trim=0 0 40 0, clip, width=.45\textwidth]{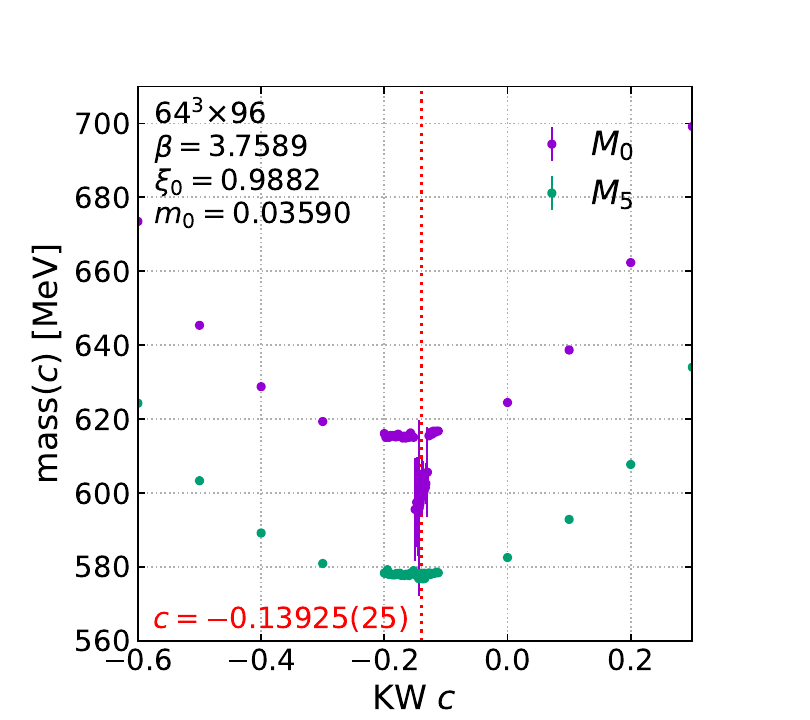} \includegraphics[width=.495\textwidth]{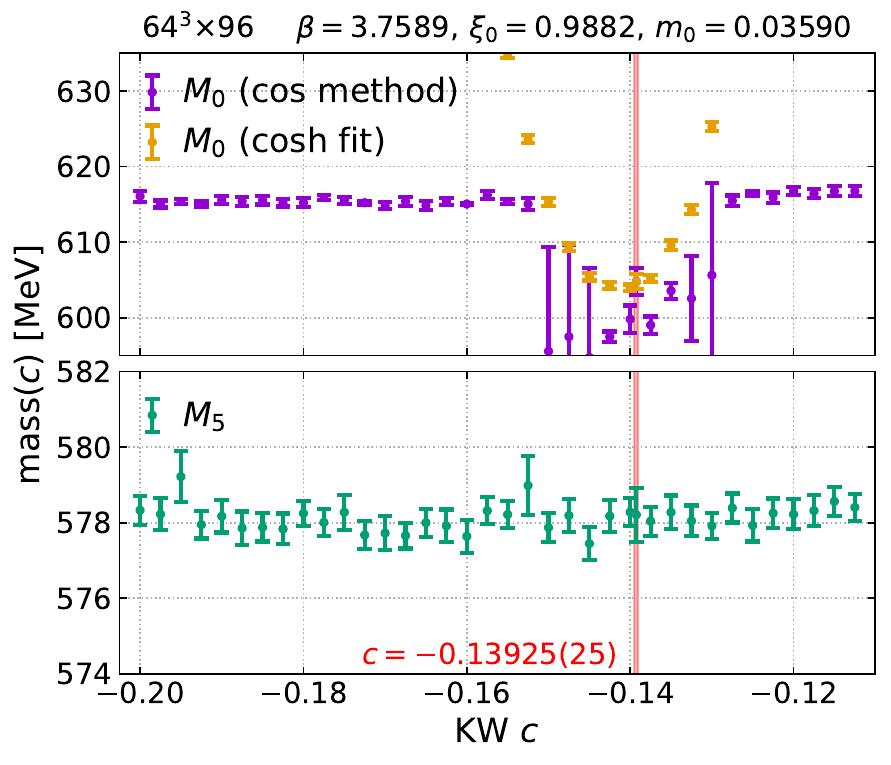} 
 \caption{(Left) Physical masses of the $\gamma_0$ and $\gamma_5$ channels as
functions of the $c$ parameter at fixed $\xi_0$ and bare mass $m_0$. $M_0(c)$
and $M_5(c)$ are concave up, with minima near the tuned value of $c$. (Right)
Zoom in near tuned $c$. While $M_5$ is very stable around tuned $c$, $M_0$
exhibits a sudden dip in the immediate vicinity of tuned $c$, which can be
observed across multiple fitting methods (explained in the text).
\label{fig:m0_m5_vs_c} } 
\end{figure} 

Until this point, we have not mentioned the bare mass parameter of the KW 
action $m_0$, which determines the ground state masses of the pseudoscalar 
$\gamma_0$ and $\gamma_5$ channels at tuned $c$ and $\xi_0$. The left plot of 
Fig. \ref{fig:m0_m5_vs_c} shows the physical $\gamma_0$ and $\gamma_5$ masses, $M_0$ and $M_5$ respectively, as functions of the $c$ parameter for fixed $\xi_0$ and 
$m_0$. We may also speak of the stability of the tuning in regard to $M_0$ and 
$M_5$. Both masses are concave up functions of $c$ with minima near the tuned 
value of $c$, indicating the stability of both quantities to small mistunings 
of $c$. An interesting effect is observed, however, in the behavior of $M_0(c)$ 
very close to the tuned $c$. While $M_5(c)$ is practically flat in the immediate 
vicinity of the tuned $c$, $M_0(c)$ experiences a sudden dip over a small interval 
around tuned $c$. This region of the plot is enlarged in the right plot of 
Fig.~\ref{fig:m0_m5_vs_c}. We observe this effect at all $\beta$ values at 
which we take measurements. Very close to the tuned $c$, the determination of $\omega_c$ is very difficult because of the finite size of the lattice.

This dip around tuned $c$ in the regime where $\omega_c$ can no longer be
reliably determined would seem to pose a great challenge to tuning c, since a
small mistuning appears to cause a significant difference in $M_0$. However,
the accuracy and the precision of the method of tuning with $\omega_c$ is such
that we are able to reliably penetrate this difficult region. The right plot of
Fig. \ref{fig:m0_m5_vs_c} shows the value of $M_0$, extracted using two 
different methods: one method utilizing a cosine fit to the beat frequency
component of the correlator, the other being a direct cosh fit to the
correlator. The latter method is only viable when the wavelength of the beat
is large compared to the lattice, hence no beat is apparently visible. The
values of $M_0$ extracted from both methods are in agreement at the tuned value
of $c$ determined from where $\omega_c=0$. We can then conclude that the
correct determination of the value of $M_0$ is robust, in that the same value 
is obtained with different fit methods, but is subordinate to a precise tuning 
of $c$ as shown in Fig.~\ref{fig:c_tuning}.

Thus, the $c$ parameter determined from the condition of vanishing beat frequency
in the $\gamma_0$ propagator is very close to the extrema of various
renormalized quantities (e.g. $\xi_f$, meson mass). The derivative of $C$-even
quantities with respect to $c$ are odd, and should vanish with a restored 
symmetry. However, the use of e.g $dM_5/dc=0$ for tuning of $c$ is hindered by
noise (see Fig.~\ref{fig:m0_m5_vs_c}).

\begin{figure}
\centering 
\includegraphics[trim=20 0 40 0, clip, width=.45\textwidth]{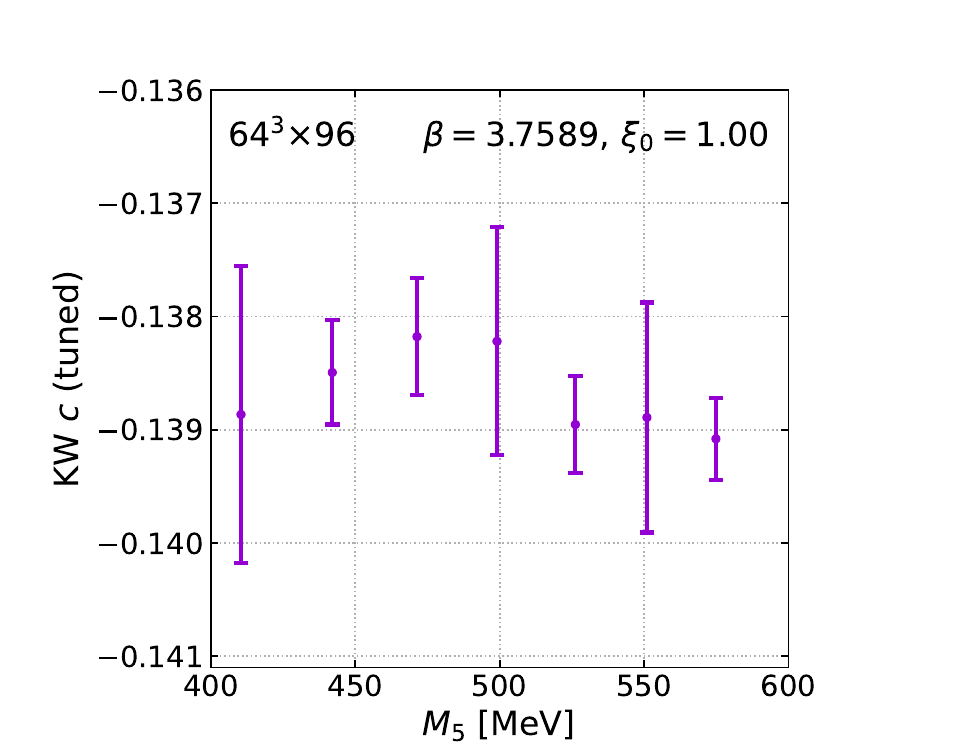} 
\includegraphics[trim=20 0 40 0, clip, width=.45\textwidth]{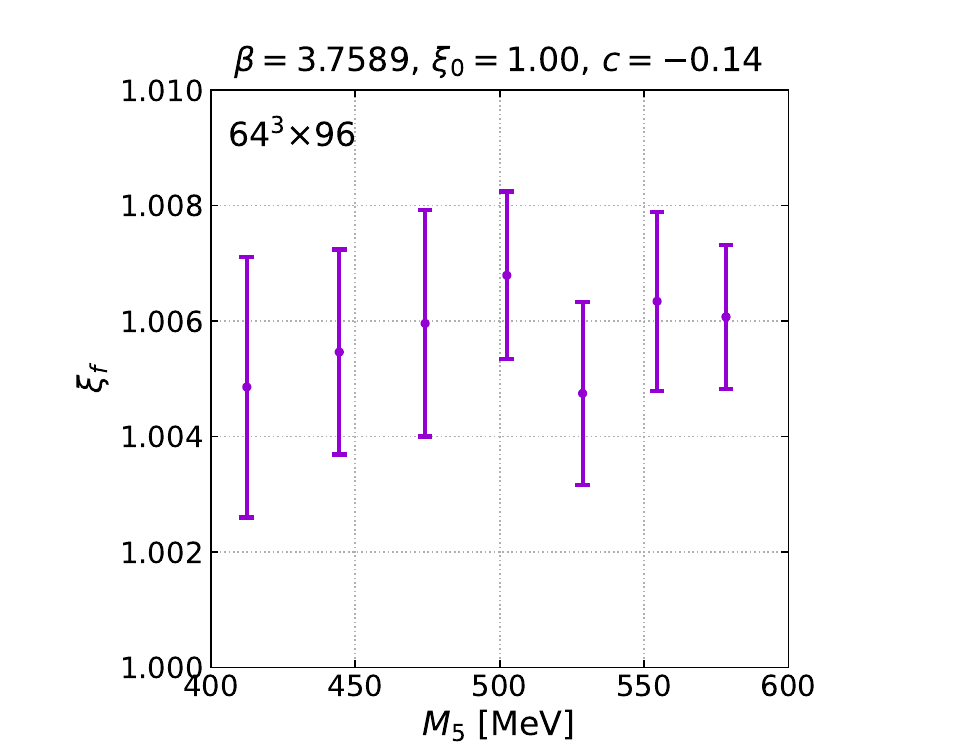} 
\caption{Stability of the tuned $c$ value at fixed $\xi_0$ (left) and the renormalized anisotropy $\xi_f$ at fixed $\xi_0$ and $c$ (right) with respect to the pseudoscalar mass $M_5$.  \label{fig:m5_variation} }
\end{figure}

\begin{figure}
 \centering 
\includegraphics[width=.45\textwidth]{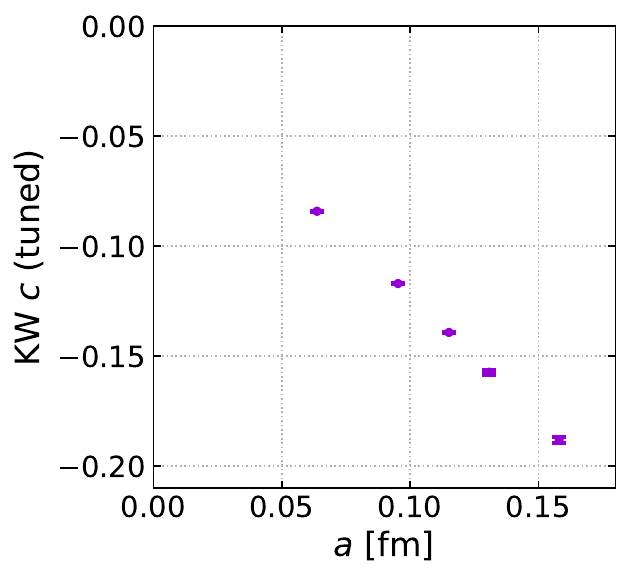}
\includegraphics[width=.45\textwidth]{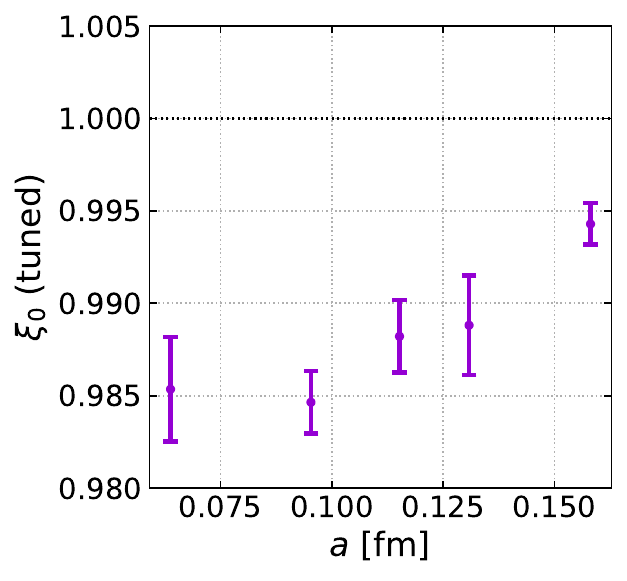}
\caption{Tuned values of the $c$ parameter (left) and $\xi_0$ (right) at various lattice spacings for constant pseudoscalar mass $M_5$. \label{fig:tuned_values} }
\end{figure}

We find that the tuned value of $c$ for a given $\xi_0$ is very stable with
respect to changes in the physical mass of the $\gamma_5$ channel $M_5$.
Further, for fixed $c$ and $\xi_0$, the renormalized fermionic anisotropy
$\xi_f$ is very stable with respect to changes in $M_5$ as well.
These results are shown in Fig. \ref{fig:m5_variation}. 
In Fig. \ref{fig:tuned_values}, we show the results of tuning $c$ and $\xi_0$ 
at several lattice spacings with the $\gamma_5$ mass held constant at 
$M_5 = 578.4$ MeV. 

A hierarchy of the bare parameters is thus established. Most critical in the 
tuning procedure is the $c$ parameter, as the physical masses of oscillating 
fermion channels are highly dependent on it. The bare anisotropy $\xi_0$ 
follows in importance. Finally, the bare mass $m_0$ (alternatively the physical 
mass $M_5$) is last, as it has the mildest effect.


\section{Taste-splitting of $\gamma_0$ and $\gamma_5$ channels} \label{taste_splitting}

\begin{figure} \centering 
\includegraphics[trim=20 0 70 0, clip,width=.45\textwidth]{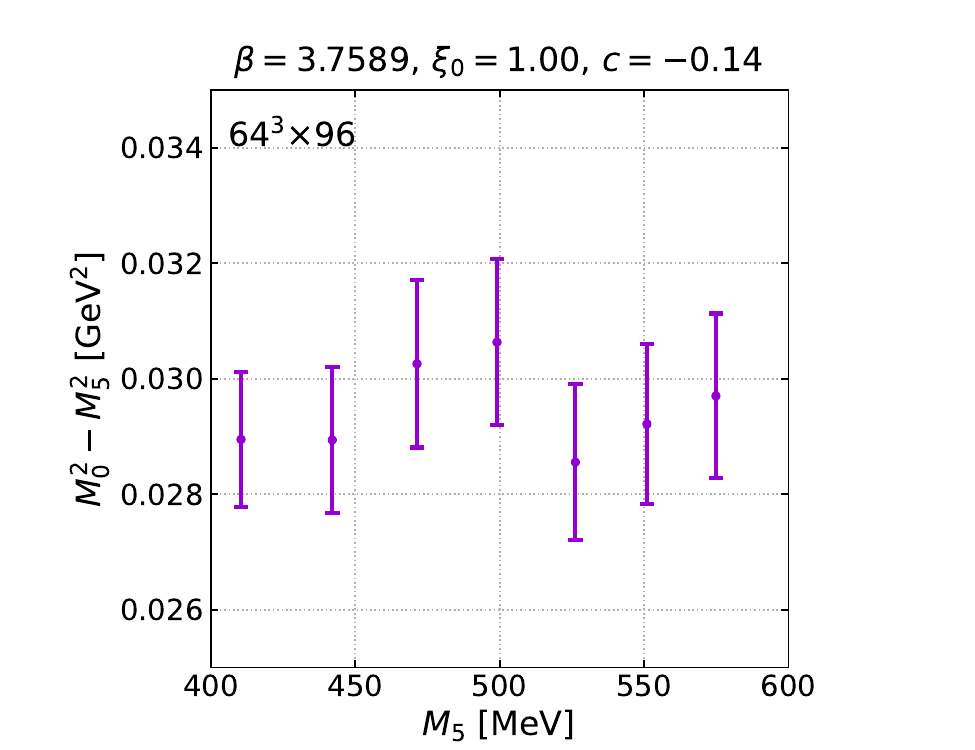}
\includegraphics[width=.45\textwidth]{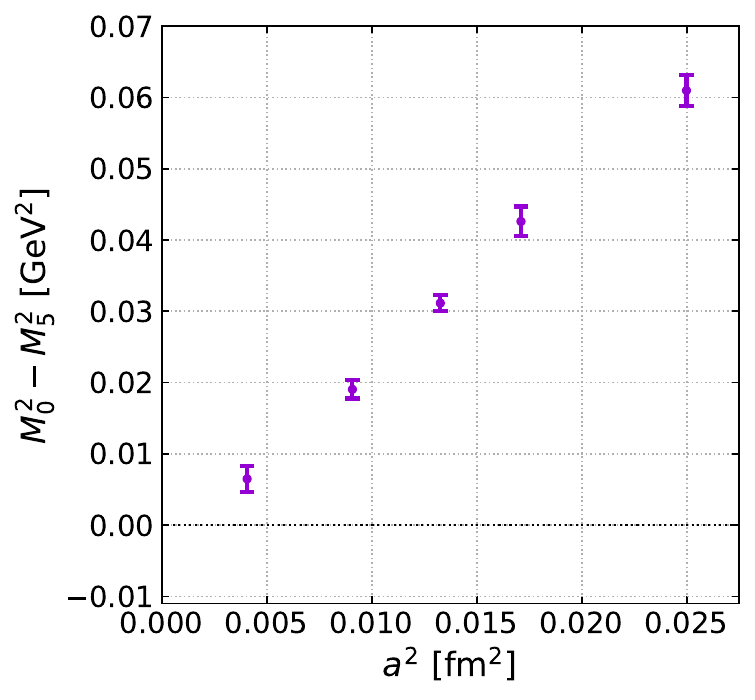}
\caption{(Left) Stability of the taste-splitting $\Delta M^2 = M_0^2 - M_5^2$
of the $\gamma_0$ and $\gamma_5$ channels with respect to the pseudoscalar
mass $M_5$.  (Right) Taste-splitting vs. lattice spacing squared.
\label{fig:taste_splitting} } 
\end{figure}

Finally, we present an investigation of the mass-splitting of the ground states
of the $\gamma_0$ an $\gamma_5$ channels, which are parity partners in the
spin-taste structure of the KW action \cite{Weber:2023kth}. The quadratic mass
difference $\Delta M^2 = M_0^2 - M_5^2$ is a quantity which is stable against
changes in physical mass $M_5$ for fixed $c$ and $\xi_0$, as shown in the left
plot of Fig. \ref{fig:taste_splitting}. In the right plot of the same figure,
we show $\Delta M^2$ at tuned $c$ and $\xi_0$ with fixed $M_5 = 578.4$ MeV, 
as a function of the lattice spacing squared. A na\"ive linear extrapolation to 
the continuum limit excluding the coarsest lattice would yield 
$\Delta M^2 < 0$. However, these lattice spacings may fall outside the linear 
scaling regime.

\section{Conclusion} \label{summary} 
We presented a mixed action study of tuning the bare parameters of the
Karsten-Wilczek action using gauge configurations generated with the staggered
4-stout fermion action. We observed the dominance of the dimension-3 KW $c$ 
parameter, which can be tuned precisely using the frequencies of oscillating 
fermionic correlation functions of the KW action. The bare anisotropy $\xi_0$ 
follows in importance, and must be tuned -- like with other anisotropic 
discretizations -- even if the target anisotropy is 1.
The physical mass of the ground state of the $\gamma_5$ channel,
dependent on the bare mass $m_0$, does not significantly effect the tuned values
of $c$ and $\xi_0$, hence tuning can be performed at a fixed pseudoscalar mass.
Lastly, we showed the mass-splitting of the parity partners $\gamma_0$ and
$\gamma_5$ at constant physical mass with tuned parameters as a function of the
lattice spacing squared.


\section*{Acknowledgments}
This work was supported by the
MKW NRW under the funding code NW21-024-A. 
R. Vig was funded by the DFG under the Project
No. 496127839. 
The authors gratefully acknowledge the Gauss Centre for Supercomputing e.V.
(\url{www.gauss-centre.eu}) for funding this project by providing computing
time on the GCS Supercomputer Juwels/Booster at Juelich Supercomputer Centre.

\end{document}